\documentclass[mathleft,fleqn,%
]{an}
\usepackage{graphicx}
\usepackage[varg]{txfonts}
\usepackage{bm}
\usepackage{natbib}
\bibpunct{(}{)}{;}{a}{}{,}
\graphicspath{{./fig/}{./png/}}
\newcommand{\uuu}{{\bm u}}

\newcommand{\AAA}{{\bm A}}

\newcommand{\BBB}{{\bm B}}

\newcommand{\JJJ}{{\bm J}}

\newcommand{\UUU}{{\bm U}}
\newcommand{\mUUU}{\overline{\bm U}}

\newcommand{\gggg}{{\bm g}}
\newcommand{\Eq}[1]{Eq.~(\ref{#1})}
\newcommand{\Equ}[1]{Equation~(\ref{#1})}

\newcommand{\Equs}[2]{Equations~(\ref{#1}) to (\ref{#2})}
\newcommand{\EQ}{\begin{equation}}
\newcommand{\EN}{\end{equation}}
\newcommand{\EQA}{\begin{eqnarray}}
\newcommand{\ENA}{\end{eqnarray}}

\newcommand{\pd}{\partial}

\newcommand{\mean}[1]{\overline{#1}}

\newcommand{\cs}{c_{\rm s}}
\newcommand{\cst}{c_{\rm s}^2}

\newcommand{\nut}{\nu_{\rm t}}

\newcommand{\urms}{u_{\rm rms}}

\newcommand{\brms}{B_{\rm rms}}

\newcommand{\Beq}{B_{\rm eq}}
\newcommand{\Ma}{{\rm Ma}}

\newcommand{\kf}{k_{\rm f}}

\newcommand{\Co}{{\rm Co}}

\newcommand{\Pm}{{\rm Pm}}

\newcommand{\Rey}{{\rm Re}}

\newcommand{\ReM}{{\rm Re}_{\rm M}}

\newcommand{\Ta}{{\rm Ta}}
\newcommand{\qij}{Q_{ij}}
\newcommand{\qxx}{Q_{xx}}
\newcommand{\qyy}{Q_{yy}}
\newcommand{\qzz}{Q_{zz}}
\newcommand{\qxy}{Q_{xy}}
\newcommand{\qxz}{Q_{xz}}
\newcommand{\qyz}{Q_{yz}}

{}

\def\onethird{{\textstyle{1\over3}}}

\def\onehalf{{\textstyle{1\over2}}}

\newcommand{\Fig}[1]{Figure~\ref{#1}} %for ApJ
\newcommand{\Figu}[1]{Figure~\ref{#1}}

\newcommand{\Table}[1]{Table~\ref{#1}}

\begin{document}

% The following seven commands are intended for editorial usage and
% should be ignored by the author(s).
\Pagespan{1}{}% Document's page range.
% If second parameter is left empty, the last page is computed
% automatically.
\Yearpublication{2019}%
\Yearsubmission{2019}%
\Month{0}%
\Volume{999}%
\Issue{0}%
\DOI{asna.201913632}%

\title{Effects of small-scale dynamo and compressibility on the $\Lambda$ effect}

\author{Petri J. K\"apyl\"a\inst{1,2}\fnmsep\thanks{Corresponding author:
        {pkaepyl@uni-goettingen.de}}
}
\titlerunning{Influence of small-scale dynamo and compressibility on the $\Lambda$ effect}
\authorrunning{P.\,J. K\"apyl\"a}
\institute{
  Institut f\"ur Astrophysik, Georg-August-Universit\"at G\"ottingen, 
  Friedrich-Hund-Platz 1, D-37077 G\"ottingen, Germany
  \and ReSoLVE Centre of Excellence, Department of Computer Science,
  Aalto University, PO Box 15400, FI-00076 Aalto, Finland
}

\received{11th Mar 2019}
\accepted{13th Aug 2019}
%\publonline{XXXX}


\abstract{The $\Lambda$ effect describes a rotation-induced
  non-diffusive contribution to the Reynolds stress. It is commonly
  held responsible for maintaining the observed differential rotation
  of the Sun and other late-type stars. Here the sensitivity of the
  $\Lambda$ effect to small-scale magnetic fields and compressibility
  is studied by means of forced turbulence simulations either with
  anisotropic forcing in fully periodic cubes or in density-stratified
  domains with isotropic forcing. Effects of small-scale magnetic
  fields are studied in cases where the magnetic fields are
  self-consistently generated by a small-scale dynamo. The results
  show that small-scale magnetic fields lead to a quenching of the
  $\Lambda$ effect which is milder than in cases where also a
  large-scale field is present. The effect of compressibility on the
  $\Lambda$ effect is negligible in the range of Mach numbers from
  0.015 to 0.8. Density stratification induces a marked anisotropy in
  the turbulence and a vertical $\Lambda$ effect if the forcing scale
  is roughly two times larger than the density scale height.}

\maketitle

\section{Introduction}

Solar and stellar differential rotation is thought to arise due to the
interaction of density-stratified convective turbulence and global
rotation of the star \citep[e.g.][]{R80,R89,RKH13}. While turbulence
is often associated with only enhanced diffusion, several
non-diffusive effects have also been discovered. Arguably the most
well-known of these in the astrophysical context is the $\alpha$
effect which leads to the generation of large-scale magnetic fields in
helical turbulence \citep{1966ZNatA..21..369S}. In the hydrodynamic
(HD) context, a non-diffusive contribution to the Reynolds stress,
also known as the $\Lambda$ effect, is thought to be crucial for the
maintenance of stellar differential rotation
\citep[e.g.][]{R80,KR93,KR95,KR05}. Considerable observational
\citep[e.g.][and references therein]{RKT14} and numerical
\citep[e.g.][]{PTBNS93,2005AN....326..315R,KB08,Kap19} evidence
support the existence of the $\Lambda$ effect in flows akin to those
in stellar convection zones. The $\Lambda$ effect occurs in rotating
anisotropic turbulence which means that angular momentum transport in
accretion disks is also likely affected by it
\citep[e.g.][]{SKKL09,KBKSN10}. Other non-diffusive HD effects include
the anisotropic kinetic alpha (AKA) effect
\citep[e.g.][]{FSS87,2018A&A...611A..15K} and the inhomogeneous
helicity effect \citep{2016PhRvE..93c3125Y}, but their role in the
maintenance of stellar differential rotation is likely to be
subdominant to the $\Lambda$ effect.

Numerical simulations of magnetohydrodynamic (MHD) convection in
spherical coordinates have reached sufficient spatial resolution that
allow the excitation of small-scale dynamo action
\citep[e.g.][]{NBBMT13,HRY14,KKOWB16}. The study of \cite{KKOWB16}
showed that differential rotation in simulations is strongly quenched
at the highest magnetic Reynolds numbers where an efficient
small-scale dynamo is excited. Furthermore, the turbulent Reynolds and
Maxwell stresses were found to have similar spatial distributions and
magnitudes but opposite signs. These findings can be interpreted as
magnetic quenching of the $\Lambda$ effect.

In a subsequent study \citep{Kap19}, the effect of large-scale
magnetic fields on the $\Lambda$ effect was studied. These results
show that the $\Lambda$ effect is significantly quenched when the
large-scale magnetic field reaches a substantial fraction of the
equipartition strength. However, imposing a large-scale field will
also induce small-scale fields due to tangling by the turbulent
motions and it is not possible to disentangle the two
contributions. Here this caveat is avoided by self-consistent
generation of small-scale magnetic fields by a small-scale dynamo in a
setup where no simultaneous large-scale dynamo is present.

While the mean-field theory of the $\Lambda$ effect is derived under
the assumption of incompressibility, the numerical simulations used to
compute the coefficients are most often fully compressible
\citep[e.g.][]{KB08,PTBNS93,KKT04}. Although the Mach numbers in these
studies are still clearly subsonic, the effects of compressibility
have not been studied in detail. Here such an effort is undertaken
with a controlled set of simulations where the minimal ingredients
(rotation and anisotropic turbulence) for the $\Lambda$ effect are
included while the Mach number is varied.

Another aspect that has not received much attention is the
contribution of density stratification to the anisotropy of turbulence
\citep[see, however][]{2012A&A...539A..35B} and the resulting
$\Lambda$ effect in rotating cases. Isolating these effects in
convection is not possible because the forcing due to the convective
instability is in itself highly anisotropic. Here this aspect is
studied with isothermal but density stratified setups where the
turbulence is driven by isotropic forcing and where the convective
instability is absent.

\section{The model}
\label{sec:model}

The model is the same as that used in \cite{KB08} and \cite{Kap19},
except in cases where gravity and density stratification are included.

\subsection{Basic equations}

Compressible HD or MHD turbulent flow in a fully periodic cube
is modeled. An isothermal equation of state with $p=\rho \cst$, where
$p$ is the pressure and $\cs$ is the constant speed of sound, is
assumed. The following set of MHD equations is solved:
\begin{eqnarray}
  \frac{\pd {\bm A}}{\pd t} &=& {\bm U}\times {\bm B} - \eta\mu_0{\bm J}, \\
  \frac{D \ln \rho}{Dt} &=& -\bm\nabla\bm\cdot{\bm U}, \label{equ:NS} \\
  \frac{D {\bm U}}{Dt} &=& \gggg -\cst \bm\nabla\ln\rho - 2\ \bm\Omega \times {\bm U} + {\bm F}^{\rm visc} + {\bm F}^{\rm force},
\end{eqnarray}
where $\AAA$ is the magnetic vector potential, $\UUU$ is the fluid
velocity, $\BBB = \bm\nabla \times \AAA$ is the magnetic field, $\JJJ
= \mu_0^{-1} \bm\nabla \times \BBB$ is the current density, $\eta$ is
the magnetic diffusivity, $\mu_0$ is the permeability of vacuum,
$\rho$ is the density, $D/Dt = \pd/\pd t -\UUU \bm{\cdot\nabla}$ is
the advective time derivative, $\gggg=(0,0,-g)$ is the acceleration
due to gravity, $\bm\Omega$ is the rotation vector, and ${\bm F}^{\rm
  visc}$ and ${\bm F}^{\rm force}$ describe the viscous force and
external forcing, respectively.

The viscous force is given by
\begin{eqnarray}
{\bm F}^{\rm visc} = \nu \left(\nabla^2\UUU + \onethird \bm\nabla\bm\nabla\bm\cdot \UUU + 2\bm{\mathsf{S}} \bm\cdot \bm\nabla\ln\rho \right),
\end{eqnarray}
where $\nu$ is the kinematic viscosity and
\begin{eqnarray}
\mathsf{S}_{ij} = \onehalf \left(\frac{\pd U_i}{\pd x_j} + \frac{\pd U_j}{\pd x_i}\right) - \onethird \delta_{ij} \frac{\pd U_k}{\pd x_k}
\end{eqnarray}
is the traceless rate of strain tensor.

The forcing term on the rhs of \Eq{equ:NS} is given by
\begin{eqnarray}
{\bm F}^{\rm force}({\bm x},t) = Re\{\bm{\mathsf{N}}\bm\cdot{\bm f}_{{\bm k}(t)} \exp[{\rm i} {\bm k}(t)\bm\cdot {\bm x} - {\rm i}\phi(t)]\},
\end{eqnarray}
where ${\bm x}=(x,y,z)$, $\bm{\mathsf{N}}= \bm{\mathsf{f}}\cs
(k\cs/\delta t)^{1/2}$ is a normalization factor, $\bm{\mathsf{f}}$
contains the dimensionless amplitudes of the forcing, $k=|{\bm k}|$,
$\delta t$ is the length of the time step, and $-\pi < \phi(t) < \pi$
is a random delta--correlated phase. The vector ${\bm f}_{\bm k}$
describes nonhelical transversal waves, with
\begin{eqnarray}
{\bm f}_{\bm k} = \frac{{\bm k} \times \hat{\bm e}}{\sqrt{ {\bm k}^2 - ({\bm k} \bm\cdot \hat{\bm e})^2 }},
\end{eqnarray}
where $\hat{\bm e}$ is an arbitrary unit vector and where the
wavenumber ${\bm k}$ is randomly chosen at every time step. The {\sc
  Pencil Code}\footnote{\url{http://github.com/pencil-code}} was used
to perform the simulations.

\subsection{Units, system parameters, and boundary conditions}

The units of length, time, density, and magnetic field are
\begin{eqnarray}
[x] = k_1^{-1}, [t]=(\cs k_1)^{-1}, [\rho] = \rho_0, [B] = \sqrt{\mu_0\rho_0} c_s,
\end{eqnarray}
where $k_1$ is the wavenumber corresponding to the scale of the domain
and $\rho_0$ is the initially uniform value of density. The forcing
amplitude $\mathsf{f}_{ij}$ is given by
\begin{eqnarray}
\mathsf{f}_{ij} = f_0 (\delta_{ij} + \delta_{iz} \cos^2 \Theta_{\bm k}
f_1/f_0),
\end{eqnarray}
where $f_0$ and $f_1$ are the amplitudes of the isotropic and
anisotropic parts, respectively, $\delta_{ij}$ is the Kronecker delta,
and $\Theta_{\bm k}$ is the angle between $\hat{\bm e}_z$ and ${\bm
  k}$. The forcing wavenumber is chosen from a narrow range around a
predefined wavenumber $\kf$. The Mach number of the flow is varied by
adjusting the sound speed $\cs$ and the forcing amplitudes $f_0$ and
$f_1$.

The rotation vector is given by
$\bm\Omega=\Omega_0(-\sin\theta,0,\cos\theta)^{\rm T}$, where $\theta$
is the angle with the vertical ($z$) direction. Viscosity and rotation
can be combined into the Taylor number
\begin{eqnarray}
\Ta = \frac{4\,\Omega_0^2 L_{\rm d}^4}{\nu^2},
\end{eqnarray}
where $L_{\rm d}=2\pi/k_1$ corresponds to the size of the
computational domain. Furthermore, in the MHD cases the magnetic
Prandtl number
\begin{eqnarray}
\Pm = \frac{\nu}{\eta},
\end{eqnarray}
is an additional system parameter. The density scale height
$H_\rho=\cst/g$ describes the stratification in cases with
$\gggg\neq0$.

In cases with $\gggg=0$ the system is fully periodic. When $\gggg \neq
0$, impenetrable stress-free boundary conditions corresponding to:
\begin{eqnarray}
U_z = \frac{\pd U_x}{\pd z}= \frac{\pd U_y}{\pd z} = 0
\end{eqnarray}
are enforced at the vertical ($z$) boundaries.

\subsection{Diagnostics quantities}

The following quantities are outcomes of the simulations that can only
be determined a posteriori. The fluid and magnetic Reynolds numbers are
given by
\begin{eqnarray}
\Rey = \frac{\urms}{\nu \kf},\ \ \  \ReM = \frac{\urms}{\eta \kf}.
\end{eqnarray}
The rotational influence on the flow is quantified by the Coriolis
number based on the forcing scale
\begin{eqnarray}
%\Omega_\star = \frac{2\,\Omega_0 \ell}{\urms},\label{equ:Co}
%PJK: superscripts elsewhere
\Omega^\star = \frac{2\,\Omega_0 \ell}{\urms},\label{equ:Co}
\end{eqnarray}
where $\ell=L_{\rm d}k_1/\kf=2\pi/\kf$. The Mach number is given by
\begin{eqnarray}
\Ma = \frac{\urms}{\cs}.\label{equ:Ma}
\end{eqnarray}
The magnetic field strength is given in terms of the equipartition
value
\begin{eqnarray}
\Beq=(\mu_0 \rho \UUU^2)^{1/2}.
\end{eqnarray}
Finally, the parameter
\begin{eqnarray}
  A_{\rm V} = \frac{\qxx+\qyy-2\qzz}{\urms^2},\label{equ:AV}
\end{eqnarray}
characterises vertical anisotropy of turbulence.

The density stratification is quantified by the ratio of the densities
at the top and bottom of the domain, $\Delta\rho=\rho_{z_{\rm
    bot}}/\rho_{z_{\rm top}}$ where $z_{\rm bot} k_1 = -\pi$ and
$z_{\rm top} k_1 = \pi$.

\begin{table}[t!]
\centering
\caption[]{Summary of the SSD simulations. Each set consists of ten
  runs where $\theta$ is varied in steps of $10\degr$. All runs have
  $\tilde{k}_{\rm f}=\kf/k_1=10$, $\Ta=6.2\cdot10^7$, $\Delta\rho=1$,
  $\cs=3$, $\Omega^\star=0.9$, $\Ma=0.05$, and $A_{\rm V}=-0.5$.}
      \label{tab:SSD}
     $$
         \begin{array}{p{0.10\linewidth}ccccccc}
           \hline
           \noalign{\smallskip}
Set   & \Pm  & f_0 & f_1/f_0 & \Rey & \ReM  & \tilde{b}_{\rm rms} & \mbox{Grid}\\ \hline 
SSD1  &  2.0  &     10^{-6}     & 4.0\cdot10^4 &  14 &     27     &  0.10  &  144^3     \\
SSD2  &  2.5  &  8\cdot10^{-4}  &       49     &  14 &     36     &  0.21  &  144^3     \\
SSD3  &  5.0  &  2\cdot10^{-3}  &       19     &  14 &     68     &  0.39  &  288^3     \\
SSD3  &   10  & 3.5\cdot10^{-3} &       10     &  14 &    135     &  0.52  &  288^3     \\
           \hline
         \end{array}
     $$
\end{table}

\begin{table}[t!]
\centering
\caption[]{Summary of the MA simulations. Grid resolution $288^3$,
  $\tilde{k}_{\rm f}=10$, $\Ta=6.2\cdot10^7$, $\Delta\rho=1$,
  $\theta=50\degr$, $\Rey=15$, $\Omega^\star=0.8$, and $A_{\rm
    V}=-0.4$.}
      \label{tab:Mach}
     $$
         \begin{array}{p{0.10\linewidth}ccccc}
           \hline
           \noalign{\smallskip}
Set   & \cs  & f_0 & f_1/f_0 & \Ma \\ \hline 
MA1  &  0.2  &     0.13      &  19 &  0.80 \\
MA2  &  0.5  &     0.03      &  19 &  0.31 \\
MA3  &   1   &     0.01      &  19 &  0.15 \\
MA4  &   2   & 4\cdot10^{-3} &  19 &  0.08 \\
MA5  &   3   & 2\cdot10^{-3} &  19 &  0.05 \\
MA6  &   5   & 9\cdot10^{-4} &  19 &  0.03 \\
MA7  &  10   & 3\cdot10^{-4} &  19 &  0.015 \\
           \hline
         \end{array}
     $$
\end{table}

\begin{table}[t!]
\centering
\caption[]{Summary of the STR simulations. Grid resolution $288^3$,
  $\cs=3$, $f_1=0$, $H_\rho k_1 =9/10$, $\Delta\rho=10^3$, and
  $\Rey=15$. The values of $\Omega^\star$ and $A_{\rm V}$ indicate the
%  extrema from the range $|z|<9\pi/10$.}
%PJK: added sentence
  extrema from the range $|z|<9\pi/10$. The starred runs were
    repeated at the same colatitudes as the SSD sets.}
      \label{tab:strat}
     $$
         \begin{array}{p{0.11\linewidth}ccccccr}
           \hline
           \noalign{\smallskip}
Run     & \tilde{k}_{\rm f} & \ell/H_\rho & f_0 [10^{-3}] &      \Ta [10^5]       & \Omega^\star   &      A_{\rm V}    \\ \hline 
$\mbox{STR1}^\star$  &        1.5        & 4.5 & 4.7      &  0.39 & 0.6 \ldots 1.3 & -0.38 \ldots 0.90 \\
STR2  &         3         & 2.2 & 4.8      &   6.2 & 0.5 \ldots 1.4 & -0.08 \ldots 0.57 \\
STR3  &         5         & 1.4 & 4.9      &    39 & 0.5 \ldots 1.2 & -0.04 \ldots 0.22 \\
$\mbox{STR4}^\star$  &        10         & 0.7 & 5.5      &  620 & 0.6 \ldots 1.0 & -0.01 \ldots 0.07 \\
           \hline
         \end{array}
     $$
\end{table}

\subsection{Data analysis}

The coefficients pertaining to the $\Lambda$ effect were extracted by
fitting the latitudinal profiles of the off-diagonal Reynolds stresses
with the same procedure as in \cite{Kap19}. The Reynolds stress is
given by $\qij=\mean{u_i u_j}$ where the overline denotes horizontal
averaging and where $\uuu=\UUU-\mUUU$ is the fluctuating velocity. In
the homogeneous cases an additional $z$-averaging is performed. The
fitting procedure assumes that the off-diagonal Reynolds stresses are
solely generated by the $\Lambda$ effect and that they can be
represented as
\begin{eqnarray}
\qxy &=& \nut \Omega_0 \mathcal{H},\\
\qyz &=& \nut \Omega_0 \mathcal{V},\\
\qxz &=& \nut \Omega_0 \mathcal{M},
\end{eqnarray}
where $\nut = \frac{2}{15} \urms \ell$ is an estimate of the turbulent
viscosity,
\begin{eqnarray}
\mathcal{H} &=& H \cos\theta,\\
\mathcal{V} &=& V \sin\theta,\\
\mathcal{M} &=& M \sin\theta \cos\theta,
\end{eqnarray}
and
\begin{eqnarray}
H &=& H^{(1)} \sin^2\theta + H^{(2)} \sin^4\theta, \label{equ:H}\\
V &=& V^{(0)} + V^{(1)} \sin^2\theta + V^{(2)} \sin^4\theta,\label{equ:V}\\
M &=& M^{(0)} + M^{(1)} \sin^2\theta + M^{(2)} \sin^4\theta.\label{equ:M}
\end{eqnarray}
The expansions can in principle contain an arbitrary number of higher
powers of $\sin^2\theta$ but here the simulations were made in such a
regime that adding higher order contributions to the coefficients does
not yield a significantly improved fit \citep[see][]{Kap19}.

Error estimates were computed by dividing the time series in three parts
and averaging over each part. The greatest deviation of these from the
average over the full data set was taken to represent the error.

\begin{figure*}[t]
\centering
\includegraphics[width=\textwidth]{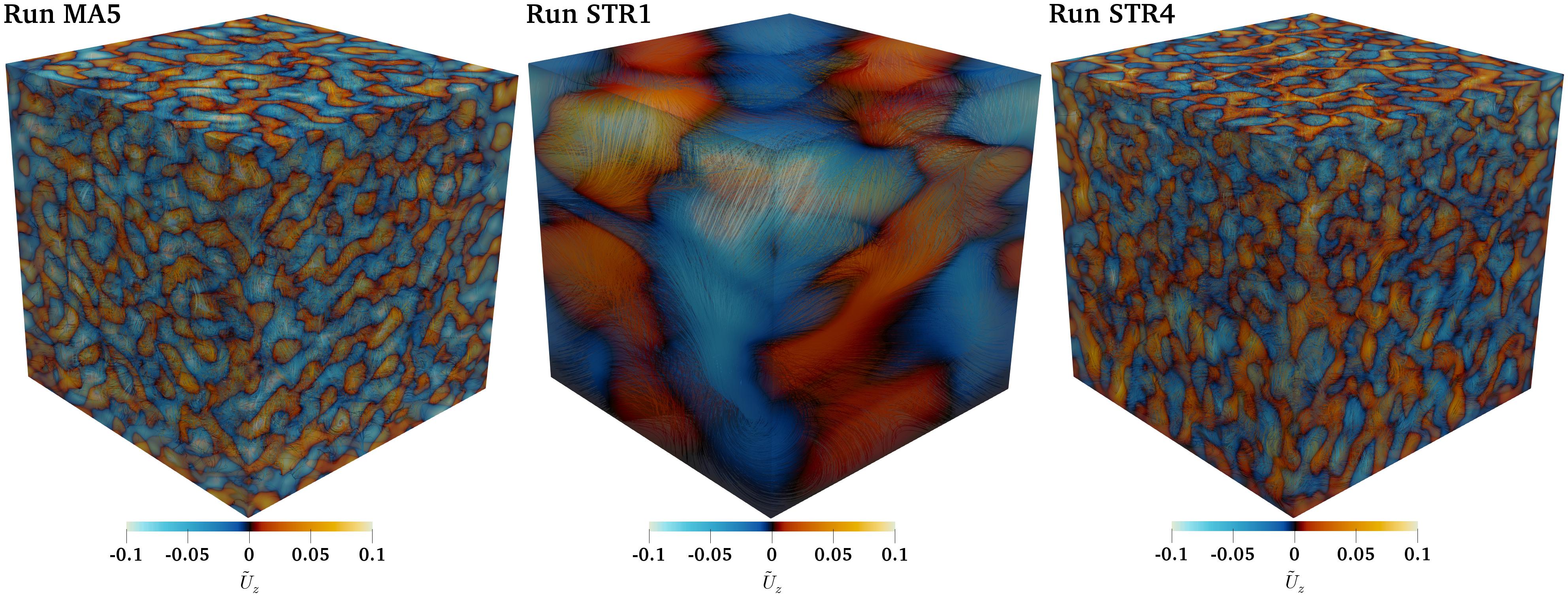}
\caption{Vertical velocity $U_z$ near the periphery of the domain and
  streamlines of the flow from Runs~MA5 (left), STR1 (middle), and
  STR4 (right).}
\label{fig:uz_boxes}
\end{figure*}

\section{Results}

Three sets of simulations were made to study the effects of
small-scale magnetic fields (Set~SSD, \Table{tab:SSD}), Mach number
(Set~MA, \Table{tab:Mach}), and density stratification (Set~STR,
\Table{tab:strat}). In Sets~SSD and MA the system is homogeneous and
anisotropically forced while in Set~STR the forcing is isotropic and a
strong density stratification is present. Visualizations of typical
flow patterns for three representative runs are shown in
\Fig{fig:uz_boxes}. A somewhat surprising result is that the presence
of strong density stratification is not clearly visible from the flow
patterns, compare the left and right panels of \Fig{fig:uz_boxes}.

\begin{figure}[t]
\centering
\includegraphics[width=\columnwidth]{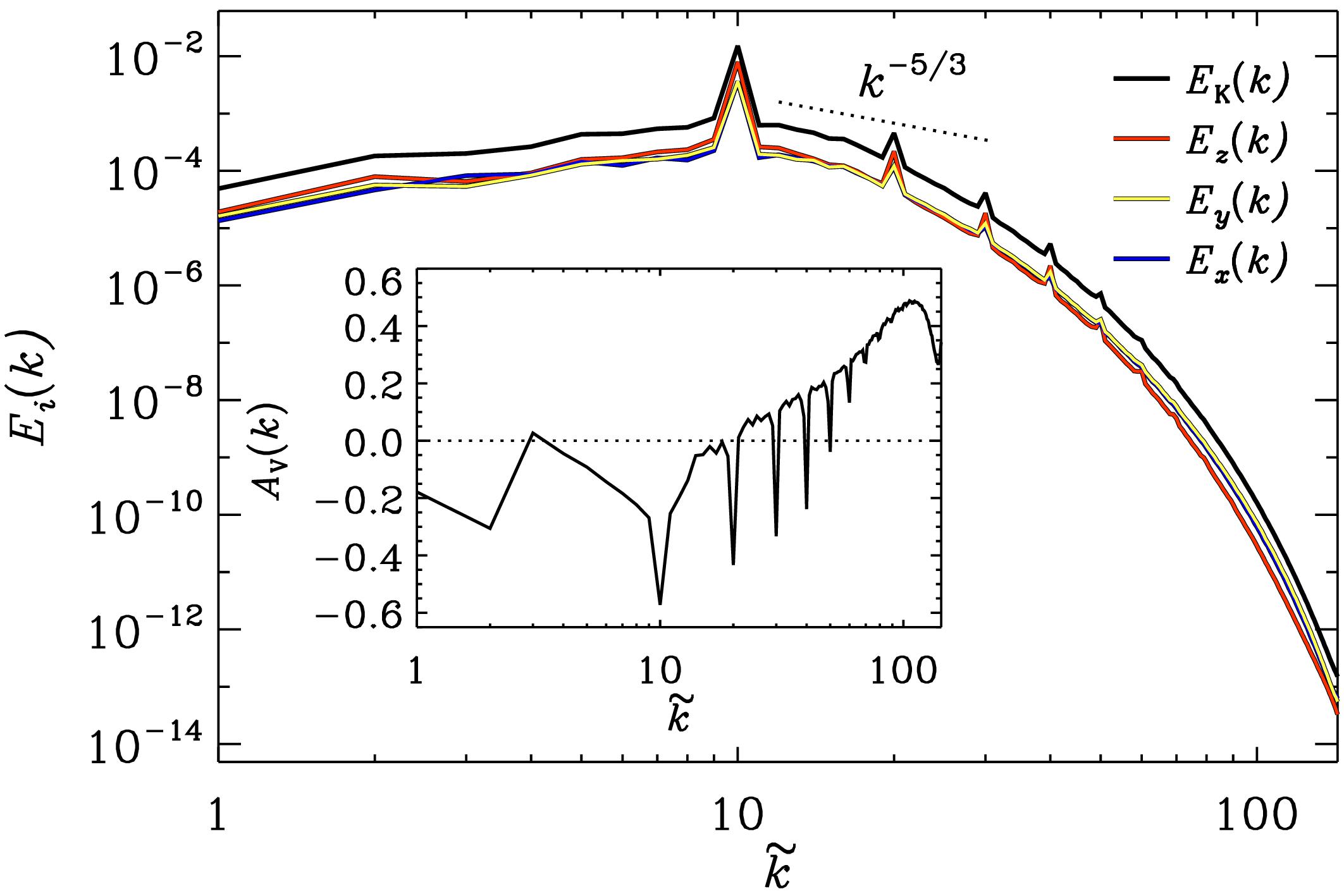}
\caption{Power spectra of the total velocity (black), and its $x$
  (blue), $y$ (yellow), and $z$ (red) components. The inset show the
  spectral anisotropy parameter $A_{\rm V}(k)$ according to
  \Eq{equ:Avk} for Run~MA7.}
\label{fig:plot_spectra}
\end{figure}

\begin{figure}[t]
\centering
\includegraphics[width=\columnwidth]{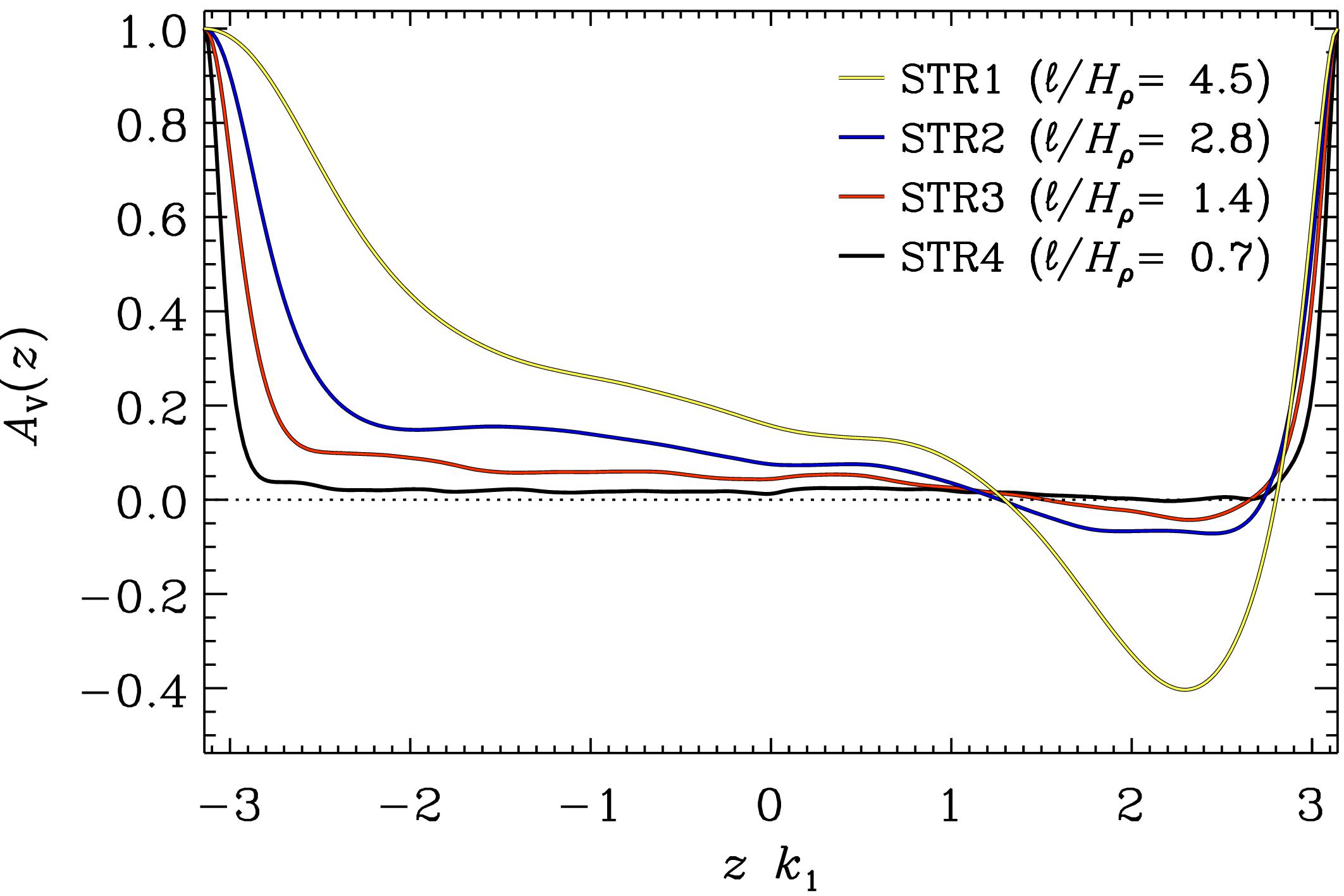}
\caption{Anisotropy parameter $A_{\rm V}(z)$ from the runs in Set~STR
  with varying $\tilde{k}_{\rm f}$.}
\label{fig:pAV}
\end{figure}

\subsection{Anisotropy of turbulence}

\Figu{fig:plot_spectra} shows the power spectra of the velocity from
Run~MA7. The power peaks near the forcing wavenumber for the total and
vertical velocity. The peaks at the overtones of the forcing
wavenumber arise because in the anisotropic case the forcing is no
longer solenoidal. The spectrum is clearly steeper than the
\cite{1941DoSSR..30..301K} (hereafter K41) $-5/3$ prediction. This is
most likely due to the insufficient scale separation between the
forcing and viscous scales which does not allow the formation of an
inertial range. Indeed, even the highest resolution simulations up to
date with $8192^3$ grid resolution are able recover only a rather
modest well-defined inertial range \citep{2017PhRvE..95b1101I}. The
viscous scale is now well resolved due to the relatively modest
Reynolds number ($\Rey\approx15$) in the current simulations. It is
also evident that the turbulence is anisotropic at all scales all the
way down to the grid scale, see the red, blue and yellow curves in
\Figu{fig:plot_spectra}. This is quantified by a spectral analogy of
the anisotropy parameter:
\begin{eqnarray}
  A_{\rm V}(k) = \frac{E_x(k)+E_y(k)-2E_z(k)}{E_{\rm K}(k)}, \label{equ:Avk}
\end{eqnarray}
where $E_{\rm K}(k)$ is the power spectrum of the total velocity and
$E_i(k)$ are the power spectra of the individual velocity
components. A representative result for $A_{\rm V}(k)$ is shown in the
inset of \Fig{fig:plot_spectra} for Run~MA7. $A_{\rm V}(k)$ has a
minimum at $k=\kf$ which is due to the fact that the forcing mainly
puts energy in the $z$ component of the velocity in this case.

While $A_{\rm V}(k)$ is mostly negative for large scales, it gradually
increases and peaks around $0.5$ for $\tilde{k}\approx100$. The fact
that the anisotropy survives to the smallest resolved scales is in
apparent disagreement with one of the cornerstones of the K41 theory
which assumes that the turbulence is fully isotropic at small enough
scales. However, the current simulations operate in a very modest
Reynolds number regime which cannot be directly compared with the K41
theory which formally applies to fully developed turbulence at very
high $\Rey$.

In addition to using explicitly anisotropic forcing, setups where
anisotropy arises naturally due to gravity and density stratification
are studied in Set~STR. By virtue of the isothermal equation of state
this setup is characterised by a constant density scale height
$H_\rho=\cst/g=9/10$ such that simulation domain contains seven scale
heights and a density contrast $\Delta\rho$ of more than a
thousand. Although a modest resolution of 288 grid points was used,
each scale height is covered by more than 40 grid points. This differs
from the case of convection where the density (and pressure) scale
height varies strongly as a function of depth and imposes much more
restrictive constraints on the grid size \citep[see, e.g.][]{KBKKR16}.

Similar setups, albeit with somewhat lower stratification, were used
in an earlier study by \cite{2012A&A...539A..35B}. They showed that
turbulence anisotropy remains small when isotropic forcing is used
unless the forcing scale is larger than the density scale height. This
is confirmed by the current simulations where the scale separation
ratio, quantified by the ratio of the forcing and system scales
$\tilde{k}_{\rm f}=\kf/k_1$, is varied between 1.5 and 10, see
\Fig{fig:pAV} and \Table{tab:strat} where a $z$-dependent variant of
\Equ{equ:AV} has been used. The density stratification-induced
anisotropy is almost non-existent in the bulk of the domain in the
case of the largest scale separation $\tilde{k}_{\rm f}=10$ or
$\ell/H_\rho=0.7$. The stress-free and impenetrable boundary
conditions enforce $U_z=0$ and lead to $A_{\rm V}=1$ at the vertical
boundaries. In the cases with poorer scale separation or larger
$\ell/H_\rho$, $A_{\rm V}$ tends to become more positive in the deep
parts and obtains negative values near the surface. For the poorest
scale separation (Run~STR1, $\ell/H_\rho=4.5$) the magnitude of the
anisotropy is comparable to typical values achieved with anisotropic
forcing in Sets~SSD and MA.

\begin{figure}[t]
\centering
\includegraphics[width=\columnwidth]{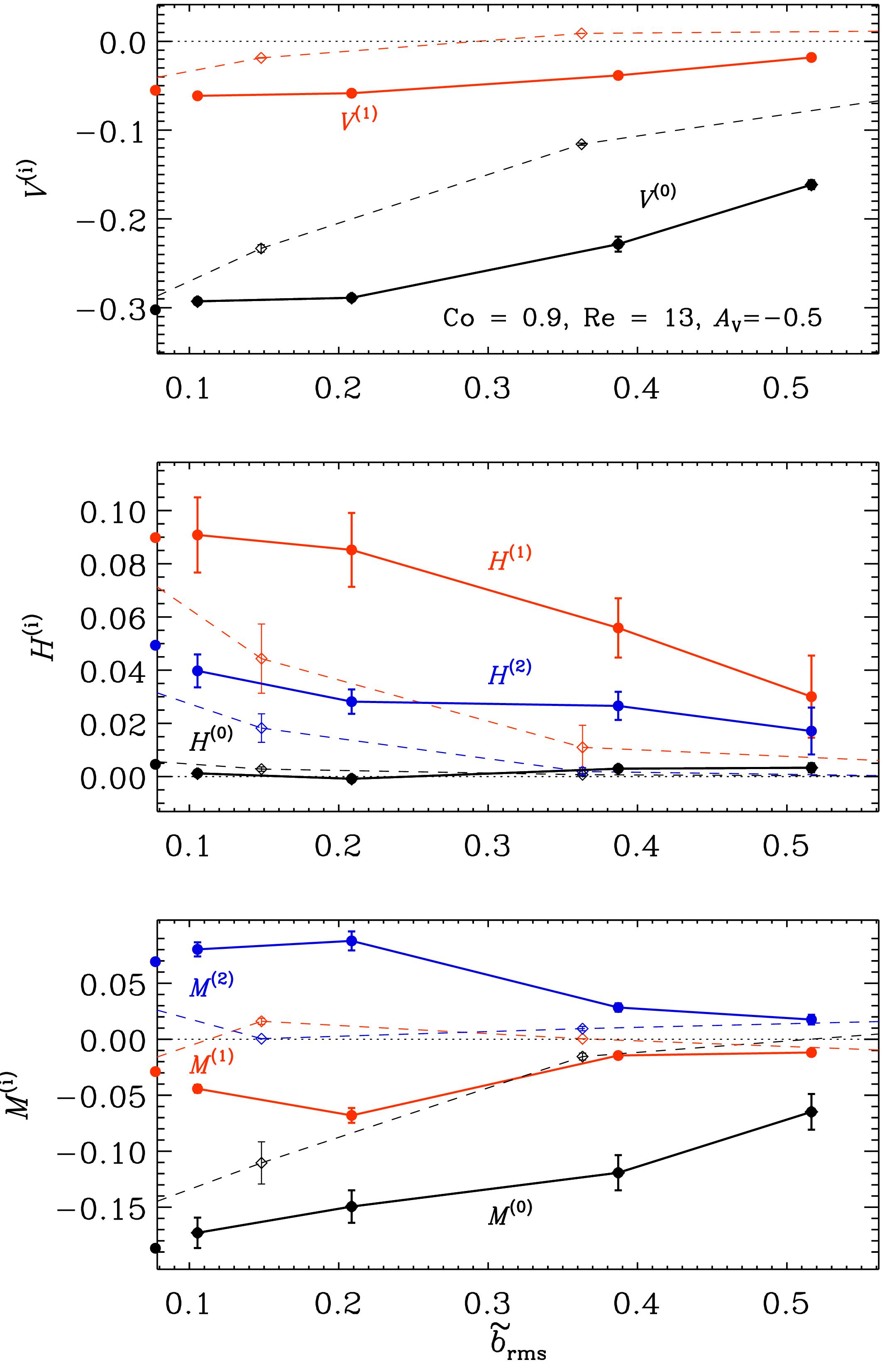}
\caption{Coefficients $V^{(i)}$ (top), $H^{(i)}$ (middle), and
  $M^{(i)}$ (bottom) as functions of the normalized magnetic field
  strength $\tilde{b}_{\rm rms}=\brms/\Beq$ from Sets~SSD1-4. The thin
  dashed lines show the corresponding quantities from runs with an
  imposed vertical field from \cite{Kap19}.}
\label{fig:pLambda_ssd}
\end{figure}

\subsection{$\Lambda$ effect}

\subsubsection{Small-scale magnetic fields}

Testing the dependence of purely small-scale magnetic fields is
possible with the current homogeneous setups in cases where the
magnetic Reynolds number exceeds that of the critical value for the
excitation of a small-scale dynamo. Due to the absence of
inhomogeneities, large-scale shear or helicity, no large-scale
magnetic fields are expected to develop. A limited range of magnetic
field strengths has been studied in the Set~SSD, see
\Table{tab:SSD}. Run~SSD1 with $\ReM=27$ corresponds to a slightly
supercritical case whereas Run~SSD4 corresponds to the highest $\ReM$
($\approx135$) that can be resolved with the adopted grid
resolution. The saturation level $\tilde{b}_{\rm rms}=\brms/\Beq$,
where $\brms$ is the rms-value of the magnetic field, increases from
roughly 10 to 50 per cent of the equipartition value in this range. In
practice the characteristics of the flow, that is the Reynolds and
Coriolis numbers and the degree of anisotropy, were kept fixed in all
runs by adjusting $f_0$ and $f_1$ while $\ReM$ was varied by changing
$\Pm$.

The results for the $\Lambda$ coefficients corresponding to
\Equs{equ:H}{equ:M} from Sets~SSD1-4 are shown in
\Fig{fig:pLambda_ssd}. All of the coefficients are quenched as the
small-scale magnetic fields increase: the values for $\tilde{b}_{\rm
  rms}\approx0.5$ are typically roughly half of their hydrodynamic
values. It is also evident that the quenching as a function of pure
small-scale fields is weaker than that in the cases where an imposed
large-scale vertical field is present, see the dashed lines in
\Fig{fig:pLambda_ssd} for the corresponding data from \cite{Kap19}.

\subsubsection{Dependence on Mach number}

Although the Mach numbers in the foregoing studies \citep{KB08,Kap19}
were typically relatively low ($\mathcal{O}(0.1)$), it cannot be ruled
out that a contribution due to compressibility is
present. Furthermore, results from low-Reynolds number shear flows
\citep{2011PhRvE..84e6314R} suggest that compressibility significantly
affects turbulent pumping for $\Ma\gtrsim0.1$.

\begin{figure}[t]
\centering
\includegraphics[width=\columnwidth]{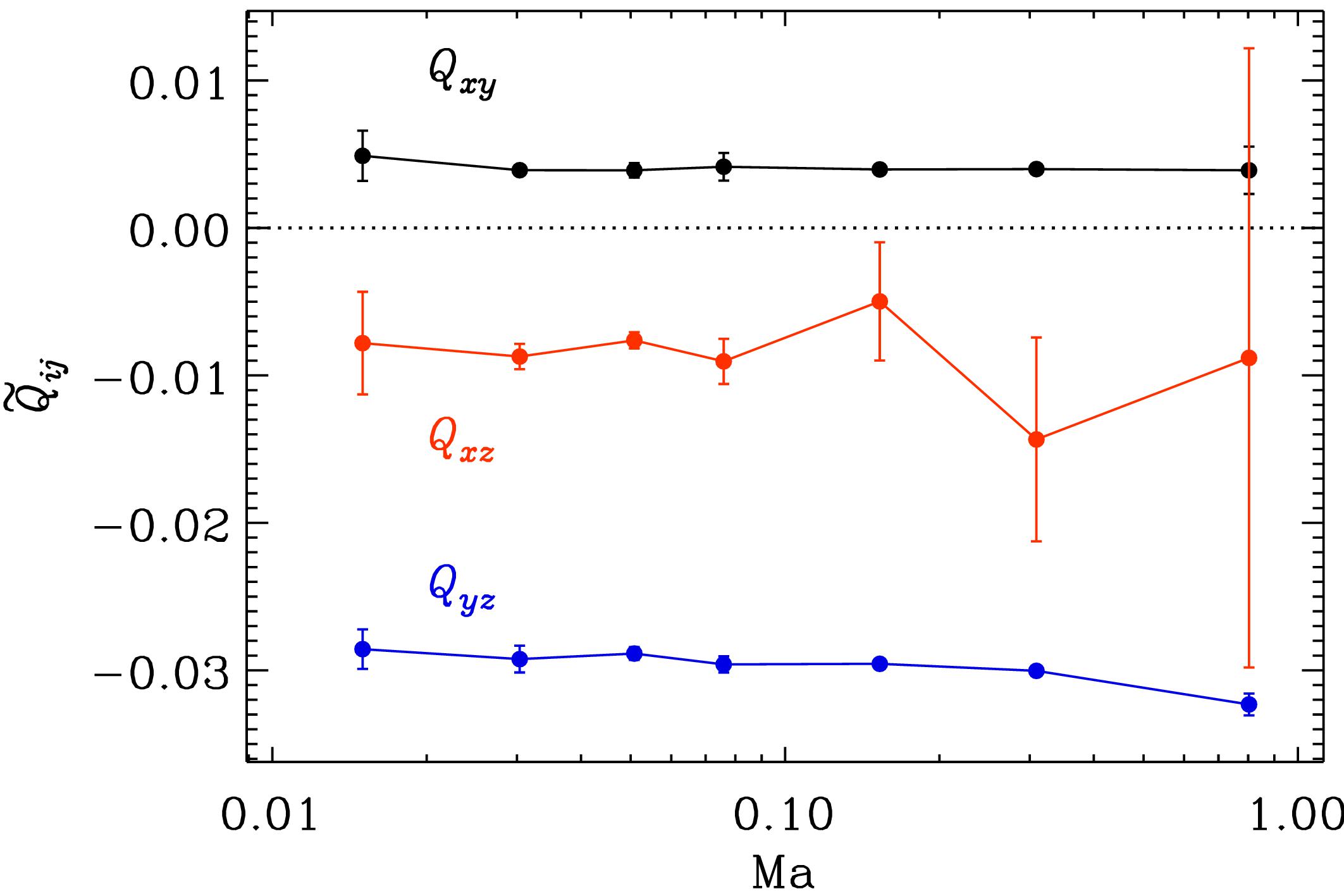}
\caption{Off-diagonal Reynolds stresses $\qxy$ (black), $\qxz$ (red),
  and $\qyz$ (blue) as functions of Mach number from Set~MA, see
  \Table{tab:Mach}.}
\label{fig:pstress_Mach}
\end{figure}

Results for the Mach number dependence from Set~MA where
$\Omega^\star=0.8$, $A_{\rm V}=-0.4$, and $\Rey=15$ are kept fixed are
shown in \Fig{fig:pstress_Mach}. The range of Mach numbers spans from
$0.015$ to $0.8$. Only the vertical stress $\qyz$ for $\Ma\approx0.8$
is statistically significantly different from the values obtained for
lower $\Ma$ and even there the change is only on the order of ten per
cent. Temporal fluctuations of $\qxz$ increase, manifested by the
drastically increased error estimates, as a function of $\Ma$ but the
time-averaged values for all runs are still consistent with a
$\Ma$-independent value.

\subsubsection{Dependence on density stratification}

In the foregoing analysis the $\Lambda$ effect resulted from the
forcing that was designed to be anisotropic. While this is the case
also in natural convection, a background density stratification also
leads to anisotropy and should hence support a $\Lambda$ effect. This
is indeed predicted by analytic theories
\citep[e.g.][]{KR05,2018ApJ...854...67P}. Here this scenario is tested
with a density-stratified setup with isotropic forcing in the Set~STR,
see \Table{tab:strat}.

The simulations considered here have $\Omega^\star\approx0.8$ which is
close to the Coriolis number where the Reynolds stresses obtain a
maximum in \cite{Kap19}. Furthermore, the simulation domain is
situated at the equator at $\theta=90\degr$ where only the vertical
Reynolds stress $\qyz$ is non-zero. In order to isolate the
contribution relevant for the $\Lambda$ effect, the horizontally
averaged horizontal mean flows $\mean{U}_x$ and $\mean{U}_y$ are
artificially removed from the solution similarly as in \cite{RKKS19}.

\begin{figure}[t]
\centering \includegraphics[width=\columnwidth]{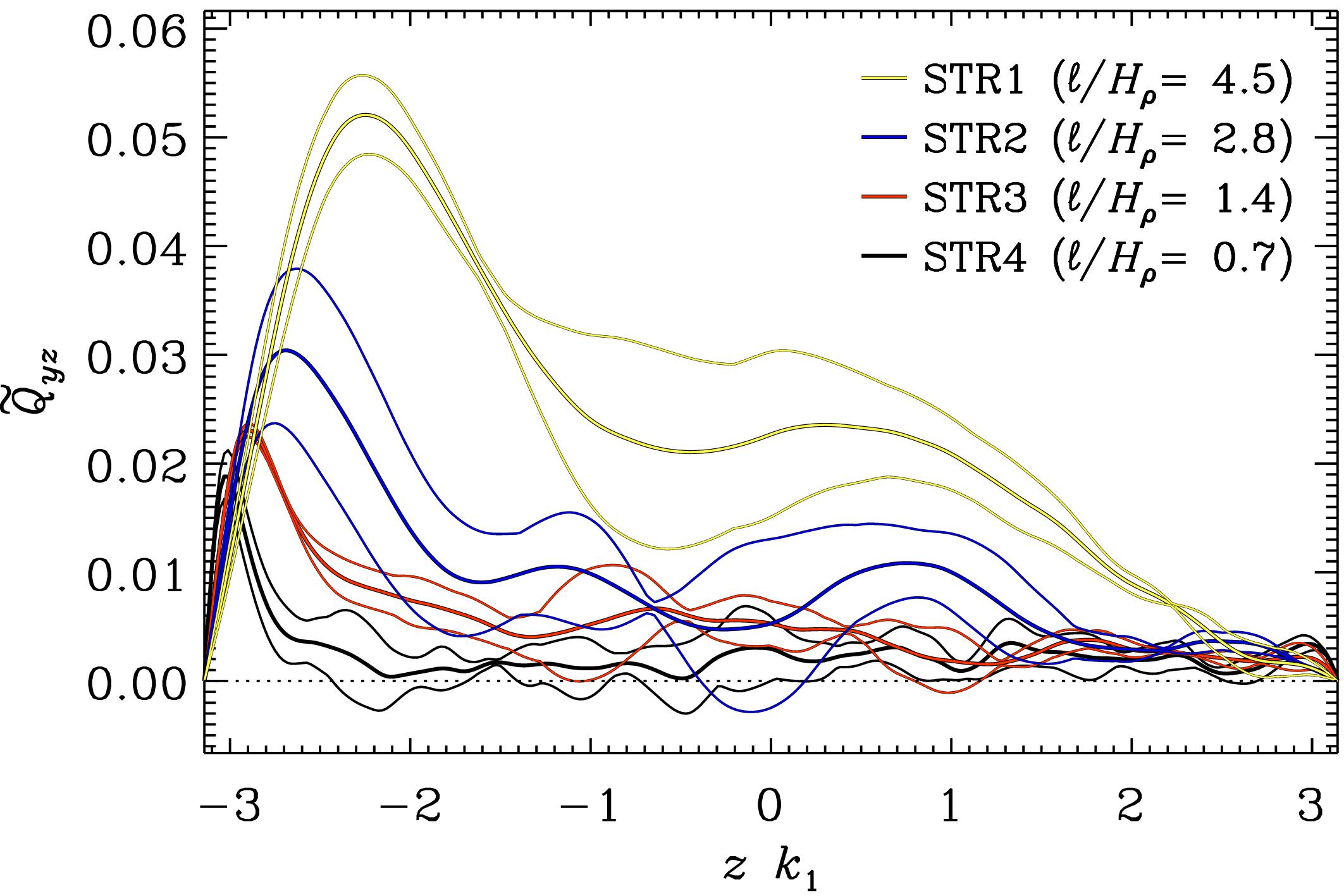}
\caption{Vertical Reynolds stress $Q_{yz}$ normalized by the squared
  rms-velocity $\urms^2(z)$ in density-stratified runs with varying
  $\ell/H_\rho$.}
\label{fig:pRyz}
\end{figure}

\begin{figure}[t]
\centering \includegraphics[width=\columnwidth]{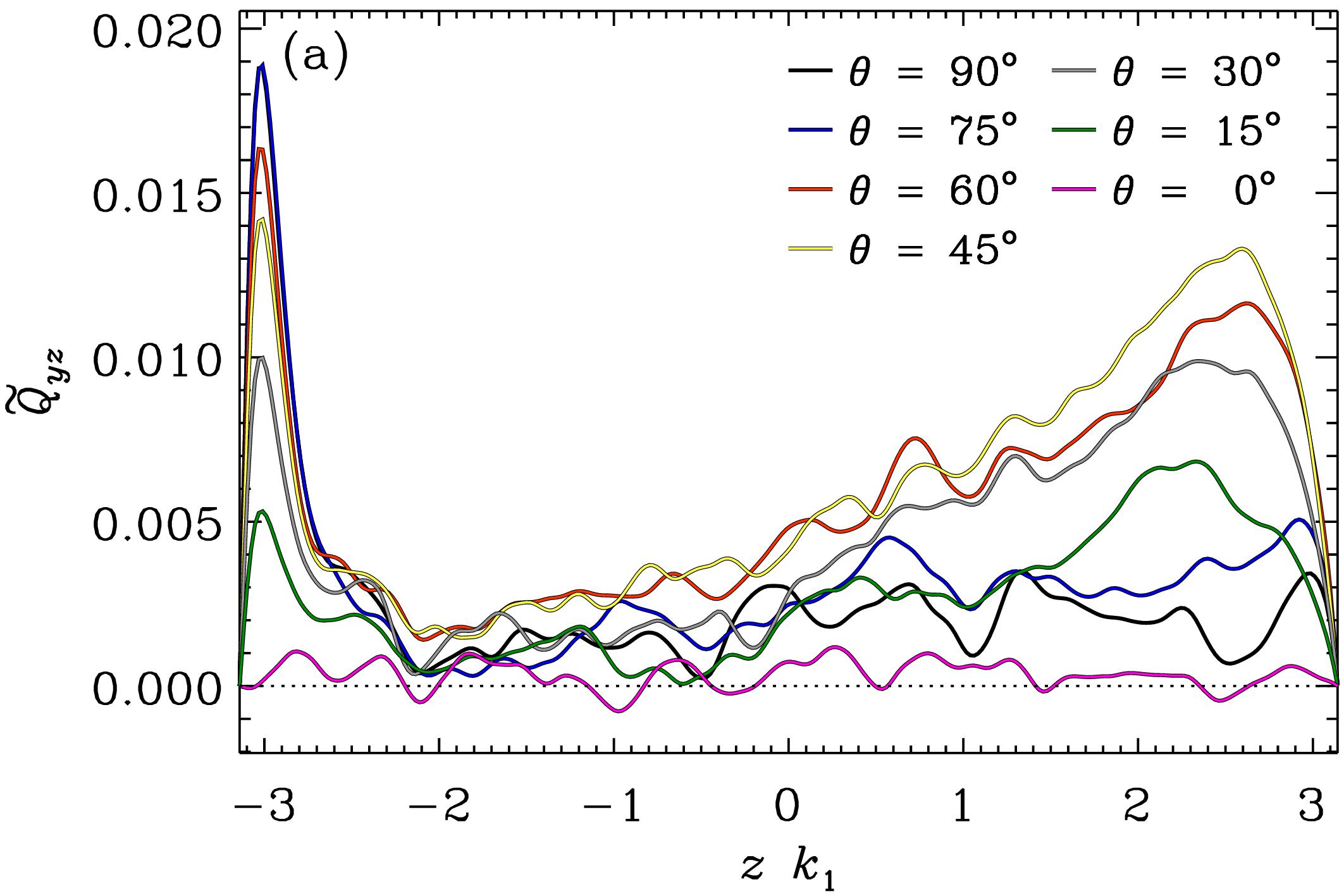}
\centering \includegraphics[width=\columnwidth]{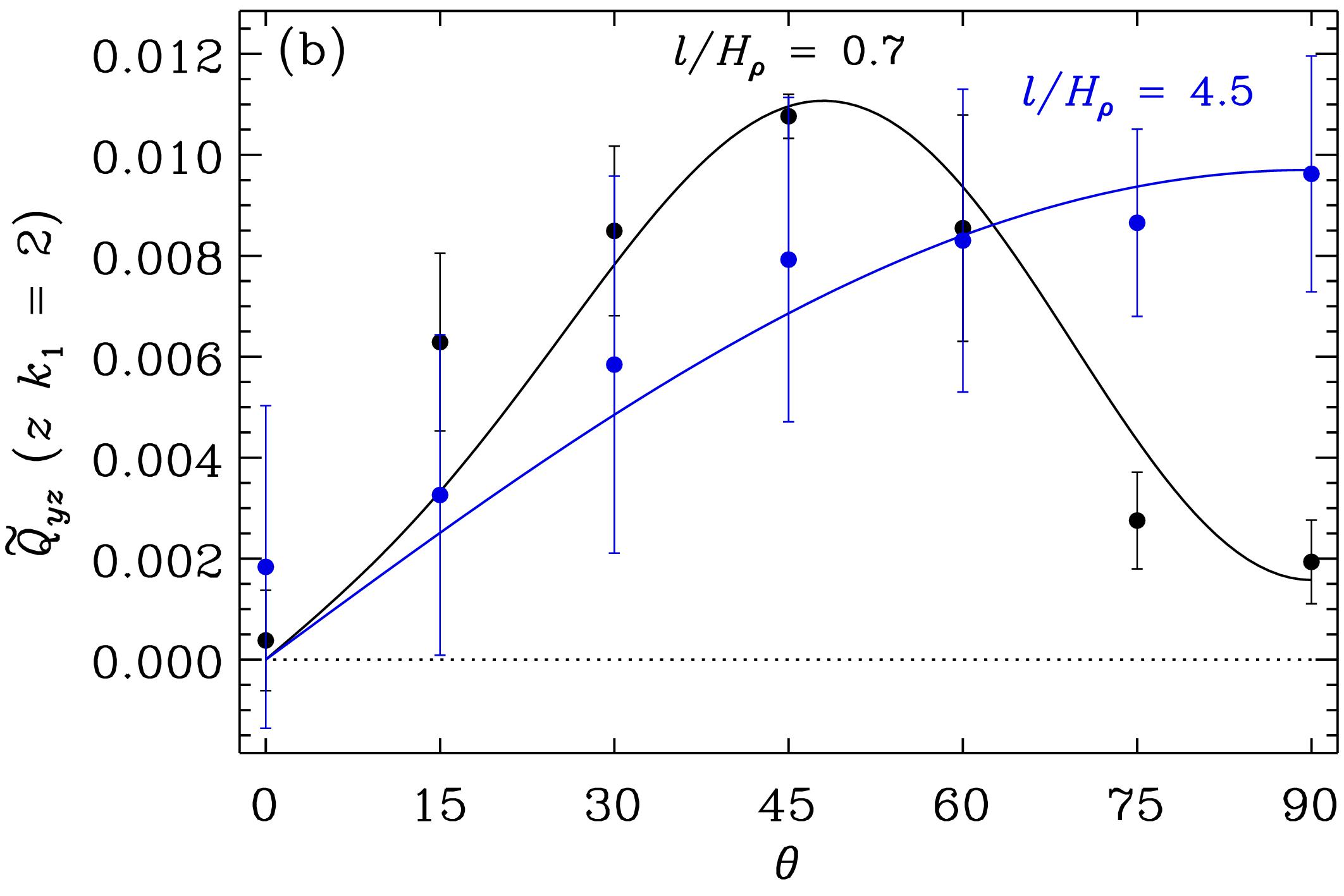}
\caption{(a) Normalized vertical Reynolds stress $Q_{yz}$ as a
    function of height from runs with $\ell/H_\rho = 0.7$ and
    $\Omega^\star=0.8$ from different colatitudes as indicated by the
    legend. (b) $\tilde{Q}_{yz}$ from $z\ k_1 =2$ as functions of
    $\theta$ from two sets with $\ell/H_\rho = 0.7$ (black) and
    $\ell/H_\rho = 4.5$ (blue). The curves show best fits according to
    \Eq{equ:V} with $V^{(0)}=0.011$, $V^{(1)}=0.027$, $V^{(2)}=-0.036$
    for $\ell/H_\rho = 0.7$ and $V^{(0)}=0.010$ for $\ell/H_\rho =
    4.5$}
\label{fig:pRyz_lat}
\end{figure}

The results for the vertical stress $\qyz$ are shown in
\Fig{fig:pRyz}. The stress is positive everywhere even in the
near-surface regions where $A_{\rm V}<0$. The results indicate that
the scale separation ratio can play an important role for the
density-induced anisotropy and consequently for the associated
$\Lambda$ effect: both are substantial in the case of large-scale
forcing and tend to approach zero when $\tilde{k}_{\rm f}$
increases. The values of $\qyz$ are of the order of a few per cent of
$\urms^2$ for $\ell/H_\rho = 4.5$ whereas for $\ell/H_\rho = 0.7$ the
effect is no longer statistically significant at the
equator. \Figu{fig:pRyz_lat}(a) shows the normalized vertical stress
$\tilde{Q}_{yz}$ from seven latitudes for the STR4 runs. The maximal
values are generally on the order of one to two per cent of the
squared rms-velocity. In this set the maximum value is obtained at
$\theta=45\degr$ and only very small values are obtained at the
equator ($\theta=90\degr$), see \Fig{fig:pRyz_lat}(b). This, however,
depends on the scale separation ratio because for a corresponding set
with $\ell/H_\rho=4.5$, $\qyz$ is consistent with a monotonic increase
toward the equator. In all of these runs the sign of $A_{\rm V}$
differs from the sign of $\qyz$. The mismatch of the signs of $\qyz$
and $A_{\rm V}$ also suggests that non-locality may play a significant
role when the scale separation ratio is small.

Another contribution to the $\Lambda$ effect due to a vertical
gradient of the Coriolis number was discussed recently by
\cite{2018ApJ...854...67P}. However, this effect is relevant only for
large Coriolis numbers and thus not applicable here. Furthermore, the
local Coriolis number $\Co=2\Omega_0 \ell/\urms(z)$ varies by a factor
between two and three in the current simulations (see the sixth column
of \Table{tab:strat}) which is mild in comparison to the variation of
four orders of magnitude in the solar convection zone as considered by
\cite{2018ApJ...854...67P}.

\section{Conclusions}

The effects of small-scale magnetic fields, compressibility and
background density stratification on the $\Lambda$ effect were studied
with numerical simulations of forced turbulence with an isothermal
equation of state.

The small-scale magnetic fields generated by a small-scale dynamo lead
to a significantly milder quenching of the $\Lambda$ effect in
comparison to cases where also a uniform large-scale field is imposed
\citep[see, e.g.][]{Kap19}. Thus is appears that the small-scale
dynamo alone could not explain the severely quenched differential
rotation in recent semi-global convection simulations
\citep{KKOWB16}. It is also conceivable that other MHD instabilities,
such as the magnetorotational instability \citep[e.g.][]{Ma11}, can be
excited in the high-resolution convection simulations, leading to 
repercussions for differential rotation.

Another aspect that has hitherto received little attention is the Mach
number dependence of the $\Lambda$ effect although most numerical
studies of the subject operate in a fully compressible regime
\citep[e.g.][]{KB08,Kap19}. The current results indicate that while
the fluctuations of the $\Lambda$ coefficients tend to increase with
$\Ma$, the mean values are consistent with a $\Ma$-independent value
at least until $\Ma\approx0.8$. Thus the effects of compressibility
have most likely not had a significant contribution to the results
regarding the $\Lambda$ effect in the previous numerical studies.

In the mean-field-theoretical treatment the $\Lambda$ effect has
distinct contributions from the anisotropy of turbulence and from
background density stratification. The former has been modeled by an
anisotropic forcing in homogeneous and fully periodic setups
\citep[e.g.][]{Kap19} while the latter requires a mean density
gradient and inevitably leads to inhomogeneity. The latter setup was
studied with a set of strongly stratified simulations where turbulence
was driven by isotropic forcing. Thus the anisotropy of the turbulence
was induced by the density stratification. The current results
indicate that the anisotropy is weak in cases where the forcing scale
is smaller or comparable with the density scale height. The vertical
velocities are suppressed (enhanced) over the horizontal components in
the deep (near-surface) parts of the simulations. The vertical
Reynolds stress and hence the $\Lambda$ effect are, however, positive
everywhere.

These results suggest an opposite sign of the vertical $\Lambda$
effect due to the density gradient in comparison to contribution from
the vertically dominated homogeneous turbulence at a comparable
Coriolis number. Further studies involving more realistic flows
(e.g.\ convection) are needed to study the relevance of these findings
for solar and stellar differential rotation.

\acknowledgements
  The computations were performed on the facilities hosted by CSC --
  IT Center for Science Ltd. in Espoo, Finland, who are administered
  by the Finnish Ministry of Education. This work was supported in part
  by the Deutsche Forschungsgemeinschaft Heisenberg programme (grant No.\
  KA 4825/1-1) and by the Academy of Finland ReSoLVE Centre of Excellence
  (grant No.\ 307411).

\bibliographystyle{an}
%\bibliography{bib_global}

\begin{thebibliography}{29}
\expandafter\ifx\csname natexlab\endcsname\relax\def\natexlab#1{#1}\fi

\bibitem[{{Brandenburg} {et~al.}(2012){Brandenburg}, {R{\"a}dler}, \&
  {Kemel}}]{2012A&A...539A..35B}
{Brandenburg}, A., {R{\"a}dler}, K.-H., \& {Kemel}, K. 2012, \aap, 539, A35

\bibitem[{{Frisch} {et~al.}(1987){Frisch}, {She}, \& {Sulem}}]{FSS87}
{Frisch}, U., {She}, Z.~S., \& {Sulem}, P.~L. 1987, Physica D Nonlinear
  Phenomena, 28, 382

\bibitem[{{Hotta} {et~al.}(2014){Hotta}, {Rempel}, \& {Yokoyama}}]{HRY14}
{Hotta}, H., {Rempel}, M., \& {Yokoyama}, T. 2014, \apj, 786, 24

\bibitem[{{Iyer} {et~al.}(2017){Iyer}, {Sreenivasan}, \&
  {Yeung}}]{2017PhRvE..95b1101I}
{Iyer}, K.~P., {Sreenivasan}, K.~R., \& {Yeung}, P.~K. 2017, \pre, 95, 021101

\bibitem[{{K{\"a}pyl{\"a}} {et~al.}(2018){K{\"a}pyl{\"a}}, {Gent},
  {V{\"a}is{\"a}l{\"a}}, \& {Sarson}}]{2018A&A...611A..15K}
{K{\"a}pyl{\"a}}, M.~J., {Gent}, F.~A., {V{\"a}is{\"a}l{\"a}}, M.~S., \&
  {Sarson}, G.~R. 2018, \aap, 611, A15

\bibitem[{{K{\"a}pyl{\"a}}(2019)}]{Kap19}
{K{\"a}pyl{\"a}}, P.~J. 2019, \aap, 622, A195

\bibitem[{{K{\"a}pyl{\"a}} \& {Brandenburg}(2008)}]{KB08}
{K{\"a}pyl{\"a}}, P.~J. \& {Brandenburg}, A. 2008, \aap, 488, 9

\bibitem[{{K{\"a}pyl{\"a}} {et~al.}(2016){K{\"a}pyl{\"a}}, {Brandenburg},
  {Kleeorin}, {K{\"a}pyl{\"a}}, \& {Rogachevskii}}]{KBKKR16}
{K{\"a}pyl{\"a}}, P.~J., {Brandenburg}, A., {Kleeorin}, N., {K{\"a}pyl{\"a}},
  M.~J., \& {Rogachevskii}, I. 2016, \aap, 588, A150

\bibitem[{{K{\"a}pyl{\"a}} {et~al.}(2010){K{\"a}pyl{\"a}}, {Brandenburg},
  {Korpi}, {Snellman}, \& {Narayan}}]{KBKSN10}
{K{\"a}pyl{\"a}}, P.~J., {Brandenburg}, A., {Korpi}, M.~J., {Snellman}, J.~E.,
  \& {Narayan}, R. 2010, \apj, 719, 67

\bibitem[{{K{\"a}pyl{\"a}} {et~al.}(2017){K{\"a}pyl{\"a}}, {K{\"a}pyl{\"a}},
  {Olspert}, {Warnecke}, \& {Brandenburg}}]{KKOWB16}
{K{\"a}pyl{\"a}}, P.~J., {K{\"a}pyl{\"a}}, M.~J., {Olspert}, N., {Warnecke},
  J., \& {Brandenburg}, A. 2017, \aap, 599, A5

\bibitem[{{K{\"a}pyl{\"a}} {et~al.}(2004){K{\"a}pyl{\"a}}, {Korpi}, \&
  {Tuominen}}]{KKT04}
{K{\"a}pyl{\"a}}, P.~J., {Korpi}, M.~J., \& {Tuominen}, I. 2004, \aap, 422, 793

\bibitem[{{Kichatinov} \& {R\"udiger}(1993)}]{KR93}
{Kichatinov}, L.~L. \& {R\"udiger}, G. 1993, \aap, 276, 96

\bibitem[{{Kitchatinov} \& {R\"udiger}(1995)}]{KR95}
{Kitchatinov}, L.~L. \& {R\"udiger}, G. 1995, \aap, 299, 446

\bibitem[{{Kitchatinov} \& {R{\"u}diger}(2005)}]{KR05}
{Kitchatinov}, L.~L. \& {R{\"u}diger}, G. 2005, Astron. Nachr., 326, 379

\bibitem[{{Kolmogorov}(1941)}]{1941DoSSR..30..301K}
{Kolmogorov}, A. 1941, Akademiia Nauk SSSR Doklady, 30, 301

\bibitem[{{Masada}(2011)}]{Ma11}
{Masada}, Y. 2011, \mnras, 411, L26

\bibitem[{{Nelson} {et~al.}(2013){Nelson}, {Brown}, {Brun}, {Miesch}, \&
  {Toomre}}]{NBBMT13}
{Nelson}, N.~J., {Brown}, B.~P., {Brun}, A.~S., {Miesch}, M.~S., \& {Toomre},
  J. 2013, \apj, 762, 73

\bibitem[{{Pipin} \& {Kosovichev}(2018)}]{2018ApJ...854...67P}
{Pipin}, V.~V. \& {Kosovichev}, A.~G. 2018, \apj, 854, 67

\bibitem[{{Pulkkinen} {et~al.}(1993){Pulkkinen}, {Tuominen}, {Brandenburg},
  {Nordlund}, \& {Stein}}]{PTBNS93}
{Pulkkinen}, P., {Tuominen}, I., {Brandenburg}, A., {Nordlund}, A., \& {Stein},
  R.~F. 1993, \aap, 267, 265

\bibitem[{{Rogachevskii} {et~al.}(2011){Rogachevskii}, {Kleeorin},
  {K{\"a}pyl{\"a}}, \& {Brandenburg}}]{2011PhRvE..84e6314R}
{Rogachevskii}, I., {Kleeorin}, N., {K{\"a}pyl{\"a}}, P.~J., \& {Brandenburg},
  A. 2011, \pre, 84, 056314

\bibitem[{{R\"udiger}(1980)}]{R80}
{R\"udiger}, G. 1980, Geophys. Astrophys. Fluid Dynam., 16, 239

\bibitem[{{R\"udiger}(1989)}]{R89}
{R\"udiger}, G. 1989, {Differential Rotation and Stellar Convection. Sun and
  Solar-type Stars} (Berlin: Akademie Verlag)

\bibitem[{{R{\"u}diger} {et~al.}(2005){R{\"u}diger}, {Egorov}, \&
  {Ziegler}}]{2005AN....326..315R}
{R{\"u}diger}, G., {Egorov}, P., \& {Ziegler}, U. 2005, Astron. Nachr., 326,
  315

\bibitem[{{R{\"u}diger} {et~al.}(2013){R{\"u}diger}, {Kitchatinov}, \&
  {Hollerbach}}]{RKH13}
{R{\"u}diger}, G., {Kitchatinov}, L.~L., \& {Hollerbach}, R. 2013, {Magnetic
  Processes in Astrophysics: theory,simulations, experiments} (Wiley-VCH)

\bibitem[{{R{\"u}diger} {et~al.}(2019){R{\"u}diger}, {K{\"u}ker},
  {K{\"a}pyl{\"a}}, \& {Strassmeier}}]{RKKS19}
{R{\"u}diger}, G., {K{\"u}ker}, M., {K{\"a}pyl{\"a}}, P.~J., \& {Strassmeier},
  K.~G. 2019, \aap, 630, A109

\bibitem[{{R{\"u}diger} {et~al.}(2014){R{\"u}diger}, {K{\"u}ker}, \&
  {Tereshin}}]{RKT14}
{R{\"u}diger}, G., {K{\"u}ker}, M., \& {Tereshin}, I. 2014, \aap, 572, L7

\bibitem[{{Snellman} {et~al.}(2009){Snellman}, {K{\"a}pyl{\"a}}, {Korpi}, \&
  {Liljestr{\"o}m}}]{SKKL09}
{Snellman}, J.~E., {K{\"a}pyl{\"a}}, P.~J., {Korpi}, M.~J., \&
  {Liljestr{\"o}m}, A.~J. 2009, \aap, 505, 955

\bibitem[{{Steenbeck} {et~al.}(1966){Steenbeck}, {Krause}, \&
  {R{\"a}dler}}]{1966ZNatA..21..369S}
{Steenbeck}, M., {Krause}, F., \& {R{\"a}dler}, K.-H. 1966, Zeitschrift
  Naturforschung Teil A, 21, 369

\bibitem[{{Yokoi} \& {Brandenburg}(2016)}]{2016PhRvE..93c3125Y}
{Yokoi}, N. \& {Brandenburg}, A. 2016, \pre, 93, 033125

\end{thebibliography}

\end{document}